\documentclass[journal]{IEEEtran}
\usepackage{cite}
\usepackage{amsmath,amssymb,amsfonts}

\usepackage{makecell}
\usepackage{algorithm}
\usepackage{algpseudocode}
\usepackage{graphicx}
\usepackage{float} 
\usepackage{subfigure}
\usepackage{textcomp}
\usepackage{xcolor}
\usepackage{hyperref}
\hypersetup{hypertex=true,
            colorlinks=true,
            linkcolor=blue,
            anchorcolor=blue,
            citecolor=blue}

\begin{document}
%
\title{Motion Planning Combines Psychological Safety and Motion Prediction for a Sense Motive Robot}

\author{Hejing~Ling,~
        Guoliang~Liu,~\IEEEmembership{Member,~IEEE,}
        and~Guohui~Tian
\thanks{Hejing Ling, Guoliang Liu and Guohui Tian are with School of Control Science and Engineering, Shandong University, Jinan, 250014 China (e-mail: 201934494@mail.sdu.edu.cn, liuguoliang@sdu.edu.cn, g.h.tian@sdu.edu.cn). (Corresponding author: Guoliang Liu.)}
\thanks{
This work was supported in part by the National Natural Science Foundation of China under Grant 91748115,
in part by the National Key Research and Development Program of China under Grant 2018YFB1306500.}
}


\maketitle

\begin{abstract}
 Human safety is the most important demand for human robot interaction and collaboration (HRIC), which not only refers to physical safety, but also includes psychological safety. Although many robots with different configurations have entered our living and working environments, the human safety problem is still an ongoing research problem in human-robot coexistence scenarios. This paper addresses the human safety issue by covering both the physical safety and psychological safety aspects. First, we introduce an adaptive robot velocity control and step size adjustment method according to human facial expressions, such that the robot can adjust its movement to keep safety when the human emotion is unusual. Second, we predict the human motion by detecting the suddenly changes of human head pose and gaze direction, such that the robot can infer whether the human attention is distracted, predict the next move of human and rebuild a repulsive force to avoid potential collision. Finally, we demonstrate our idea using a 7 DOF TIAGo robot in a dynamic HRIC environment, which shows that the robot becomes sense motive, and responds to human action and emotion changes quickly and efficiently. 
\end{abstract}


\IEEEpeerreviewmaketitle
\section{Introduction}

\IEEEPARstart{I}n the past few decades, industrial robots have been widely used in isolated workspaces considering worker safety. With the booming development of robotics and artificial intelligence, the research and design of interactive and collaborative robots have attracted the attention of many researchers \cite{1545389} \cite{yi2019mobile}. What follows is the safety problem of human-robot coexistence, due to the uncertainty of human actions, robots need to avoid human in a timely and move properly to avoid potential harm to human. To ensure the physical safety of workers, people usually set up safe workspace for industrial robots and prohibited workers from approaching them. Nowadays, human and robot need to coexist in a shared workspace, so it is necessary to track people's motion in real-time, and research on efficient and stable collision avoidance strategies for robot. Furthermore, in order to improve the social experience of human robot interaction, it is necessary to ensure people’s psychological safety. As mentioned in the paper \cite{herrero2020influence} \cite{bhandari2016emotional}, psychological states have a greater impact on our perception of risk and the behaviours we will take. Therefore, we must fully consider human psychological and optimize the control of robot to achieve better HRIC \cite{lasota2017survey} \cite{weng2009toward}. 

\begin{figure}[!tb]
\centerline{
\includegraphics[width=0.99\linewidth]{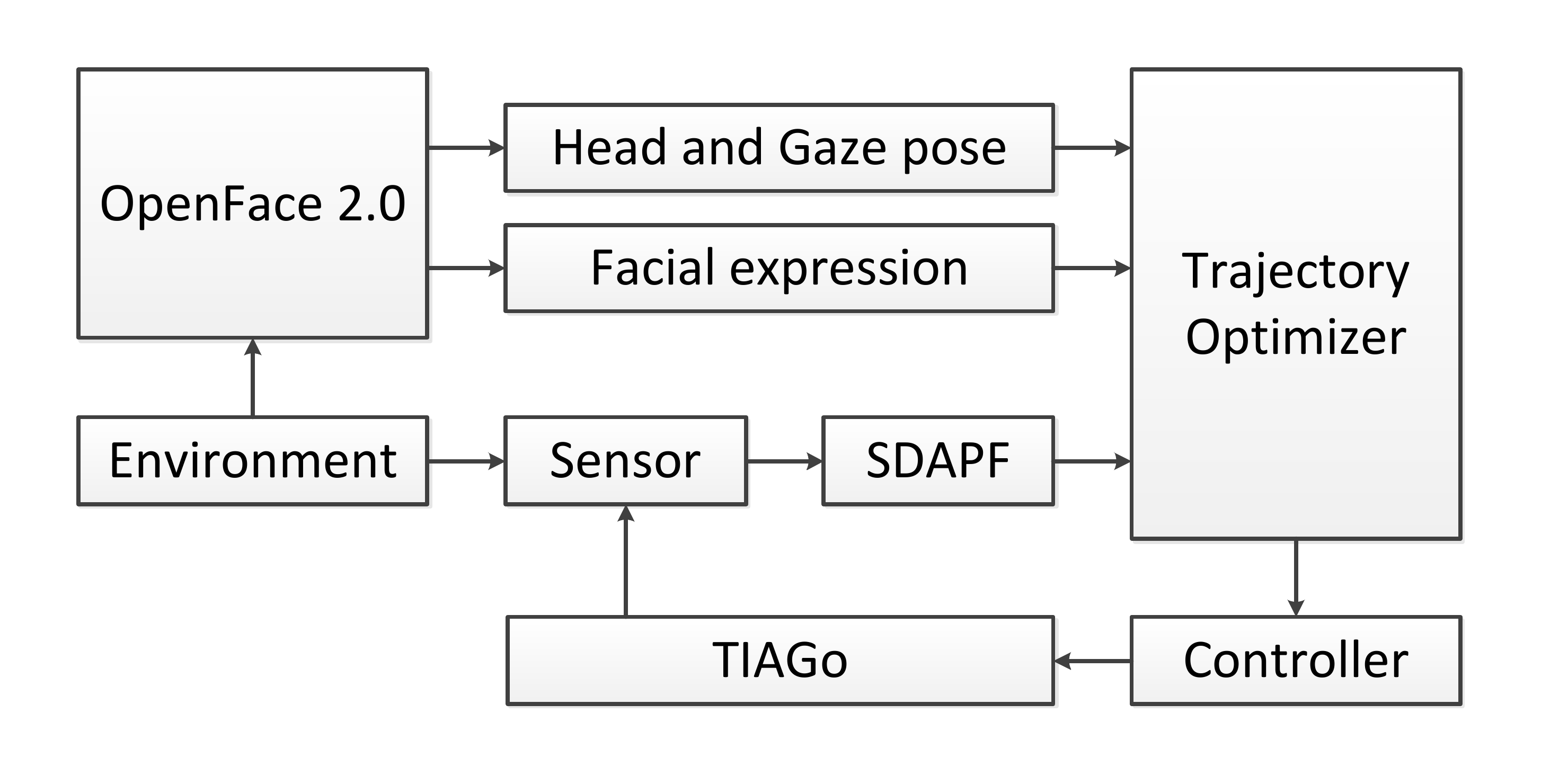}}
    \caption{Overall process of our motion planning system considering human psychological safety and motion prediction based on TIAGo platform.}
\label{fig:1}
\end{figure}

In this paper, we propose an advanced motion planning algorithm that uses facial expressions, head pose and gaze angle to estimate human psychology states and predict future move direction, and an optimized control strategy of SDAPF (sampling and danger-index based artificial field) \cite{liu2020online} is proposed for physical safety and psychological safety in HRIC scenarios. Our approach is convenient and real-time, and does not require to set up complicated environment. Specifically, we use OpenFace 2.0 \cite{baltrusaitis2018openface} to detect the facial action units (AUs) of the person in each frame of the camera, and then classify human psychology states according to the different combination of the AUs, which can be used to control the velocity and step size of the robot. In addition, we predict the human motion according to the head pose and gaze angle outputted by OpenFace, such that an improved repulsive force is proposed to avoid potential collision and optimize the robot trajectory in an uncertain environment. The overall diagram of our motion planing system for a sense motive robot is shown in Fig. \ref{fig:1}. The main advantages of our approach include: 

\begin{itemize}
\item To ensure human psychological safety, we propose an adaptive robot velocity and step size control method by using a convenient and real-time detection of human facial expressions.
\item Our method can optimize the repulsive force function to avoid potential collisions by predicting the movement of the human based on the head pose and gaze angle information.
\end{itemize}

We verify the effectiveness of the safe interaction between the 7-DOF TIAGo robot and moving human arm, and prove the advantages of our dynamic collision avoidance algorithm considering human physical and psychological safety. The rest of paper is organized as follows. In section \uppercase\expandafter{\romannumeral2}, we conduct a brief survey of previous work. Section \uppercase\expandafter{\romannumeral3} describes how we use OpenFace 2.0 to get facial expressions, head pose and gaze angle. How to optimize the collision avoidance algorithm based on the data from OpenFace 2.0 is described in section \uppercase\expandafter{\romannumeral4}. The experiment results of our approach are analyzed in section \uppercase\expandafter{\romannumeral5}. Finally, we summarize our work in section \uppercase\expandafter{\romannumeral6}.

\section{Related Works}

\subsection{Facial Expression Recognition}

Facial expressions contain richful human behavior information, which is a form of expression of human psychology. In daily life, people can fully and subtly express their thoughts and feelings through facial expressions, and they can also distinguish other's mental activities through other's facial expressions. Therefore, facial expression recognition has attracted a large number of researchers. There are different methods to recognize emotion such as electroencephalogram (EEG), Galvanic Skin Response (GSR), speech analysis, facial expressions, visual scanning behavior. However, with the popularity of deep learning, facial expression recognition based on images has made great progress.

Traditional facial expression recognition methods are based on Support Vector Machine (SVM) \cite{fu2013binary} and Hidden Markov Model (HMM) \cite{bansal2012novel}. First, people manually design features such as Gabor, LGBP and HOG that are used to extract the appearance features from images, and then classifiers are used for facial expression classification,  such as SVM or Adaboost. Feature extraction and classification are two separate processes and cannot be integrated into an end-to-end model. Zhang et al. \cite{zhang2016study} use the SVM-based method to carry out experiments on emotion recognition. The experiment show that SVM performs poorly on multi-emotion classification problems and is sensitive to the choice of kernel function.

In recent years, neural networks have been used for facial expression recognition. Ozdemir et al. \cite{ozdemir2019real} use CNN-based LeNet architecture to recognize emotion by merging three different datasets (KDEF, JAFFE and their custom dataset) and then training LeNet to obtain higher accuracy for emotion recognition. Chang et al. \cite{chang2018facial} presented a simple and efficient CNN model to extract facial features, and proposed a complexity perception classification (CPC) algorithm for facial expression recognition. The CPC algorithm divided the dataset into a simple sample subspace and a complex sample subspace by evaluating the complexity of facial features that are suitable for classification.

\subsection{Motion Planning}
For safe and efficient HRI, ensuring human safety is of paramount importance. Therefore, robots are required to be able to perform motion planning smoothly in real-time. The traditional artificial potential field (APF) method \cite{khatib1986real} designs a virtual force field in the environment, and the target point generates an attractive force $F_{att}$ to the mobile robot. Obstacles generate repulsive force $F_{rep}$ to the robot. Finally, the resultant force $F$ of attractive and repulsive force is used to control the movement of the robot. APF method is simple, practical, real-time and convenient for real-time control of the robot controller. It is widely used in real-time obstacle avoidance and smooth control. However, there are defects such as local minimum and goal nonreachable with obstacle nearby (GNRON) problem. Therefore, there are many improved APF methods. For the defects that attractive force is too large due to robot far away from goal and GNRON problem, some new repulsive or attractive force functions are introduced \cite{shi2007study,du2019real,sun2019smart, liu2020online}.

The dynamic window approach (DWA) \cite{fox1997dynamic} is a velocity-based local planner, which mainly samples multiple velocities in the velocity space, and simulates the robot's trajectory at these velocities for a certain period of time, and obtains multiple sets of trajectories. The trajectories are then evaluated, and the velocity corresponding to the optimal trajectory is selected to drive the robot. Rapidly exploring random tree (RRT) \cite{lavalle1998rapidly} using collision detection of sampling points in the space, avoiding modelling for the space, can explore the space faster and more effective, can effectively solve the path planning problems of high-dimensional space and complex constraints, which is suitable for solving the path planning of multi-degree-of-freedom robot in complex and dynamic environments. But one disadvantage of RRT is that it is difficult to find a path in an environment with narrow space. Because the area of the narrow space is small, the probability of being occupied by robots is low. In this case, the convergence speed of RRT is slow and the efficiency will be greatly reduced. Pan et al. \cite{pan2016fast} used the results of previous collision detection to predict the collision probability of new sampling points, and then improved the performance of sampling-based motion planning by combining probabilistic collision checking methods with probabilistic roadmaps (PRM) and RRT.

In certain HRI scenarios, robot cannot plan an appropriate trajectory or does not have enough time to replan, even the movement of the human may conflict with the initial planning trajectory of robot, such that it is difficult to ensure the safety of human in a complex dynamic environment. Therefore, predicting human actions is very important in a dynamic HRI environment. Kanazawa et al. \cite{kanazawa2019adaptive} determine the current task and compute the next position of the robot by modelling human movement. Park et al. \cite{park2017intention} use offline learning of human actions along with temporal coherence to predict the human actions, and use Gaussian distribution to predict human motion and calculate collision probability for safe motion planning.

For safe HRI, maintaining physical safety is the focus of the research and industrial community, but ensuring psychological safety is also crucial. Maintaining psychological safety includes that humans believe that the interaction with the robot is safe and will not cause any psychological discomfort to them due to the robot's velocity and trajectory. By simply preventing impending collisions to maintain physical safety can lead to a discomfort for human psychology \cite{lasota2015analyzing}. Psychological safety needs to be ensured in HRI, which means the velocity and trajectory of robot can be adjusted by detecting the human facial expressions, head pose and gaze angle. Yamamoto et al. \cite{yamamoto1999avoidance} consider emotions in robot control and verify the effectiveness of robot control with emotion recognition. Williams et al. \cite{williams2015emotion} use the learning classifier system (LCS) to analyze emotions for better navigation, i.e., navigation system considers emotion reduces the total number of collisions and shortens the navigation time compared with the non-adaptive navigation system .

\section{Facial behavior analysis}

\subsection{A Facial Behavior Analysis Tool: OpenFace 2.0}

\begin{figure}[!t]
\centering
\subfigure[The estimation of facial landmarks, head pose and gaze angle.]{
\includegraphics[width=3.7cm]{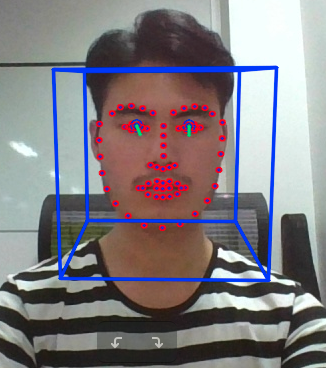}
}
\quad
\subfigure[Facial action units (AUs).]{
\includegraphics[width=4.0cm]{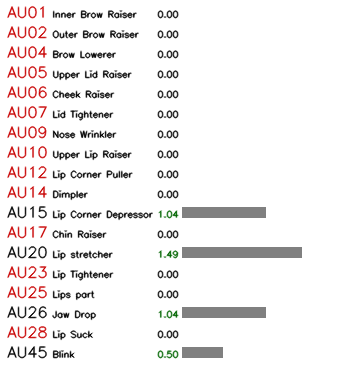}
}
\caption{The estimation results of facial landmarks, head pose and gaze angle of a sample image from a camera.}
\label{fig:3}
\end{figure}

There are many ways to detect emotional states, such as EEG and GSR. However, these methods require to set up a complex experimental environment in order to achieve effective experimental results. In contrast, this paper tries to simulate the human social interactions, and uses facial expressions to infer emotional states to ensure psychological safety. As for facial expression recognition, OpenFace 2.0 is one of the most complete and easy-to-use tool, which is capable of accurate facial landmark detection, head pose estimation, facial action units (AUs) recognition, and eye-gaze estimation, as shown in Fig. \ref{fig:3}.

The estimated head pose and gaze angle are represented as two vectors: head pose vector and gaze vector. The first three elements of head pose vector refers to X, Y, Z components of the distance between the head and camera. The rest three elements of head pose vector are pitch, yaw, roll angles. The gaze-angle is composed of two items, which represent gaze orientation for left-right and up-down directions respectively. 

OpenFace 2.0 recognizes AUs with optimized linear kernel Support Vector Machines using person specific normalization and prediction correction. Experimental results demonstrate that OpenFace 2.0 has a better performance and a distinct speed advantage than recent deep learning methods.

\subsection{Facial Expression Recognition}

In order to analyze the emotional states, it is necessary to obtain facial expressions based on AUs. Wikipedia gives a combination of AUs for six general expression states (including anger, disgust, fear, happy, sadness and surprise) according to Facial Action Coding System (FACS), which plays a very prodigious role in the field of face recognition. FACS is proposed for the purpose of finding a list of the muscles which can response separately according to changes of facial appearance. Therefore, we can analyze facial actions on the basis of the AUs (the result of muscular action). However, each person's appearance is different, so it is difficult to generalize. Although OpenFace 2.0 takes into account personal differences in AUs detection, pre-experiments are still needed to get the most obvious action units for different facial expressions of each person, which helps to get stable and reliable facial expression recognition. Considering the similarity of emotions and relevance to the HRIC safety, we only consider three facial expressions that can affect HRIC: happy, sad and surprise. 

\begin{table}[!t]
	\caption{The correspondence between facial expressions and AUs.}
	\centering
	\renewcommand\arraystretch{2}
	\setlength{\tabcolsep}{6.0mm}{
	\begin{tabular}{|c|c|}
		\hline 
			\textbf{Facial Expressions}&\textbf{AUs}\\
		\hline
			\text{Happy}&\text{AU06 + AU07 + AU12}\\
		\hline
			\text{Sad}&\text{AU04 + AU15 + AU17}\\
		\hline
			\text{Surprise}&\text{AU01 + AU02 + AU25}\\
		\hline
	\end{tabular}
	}
	\label{tab:my_label}
\end{table}

\begin{figure}[!t]
	\centerline{
	\includegraphics[width=1.0\linewidth]{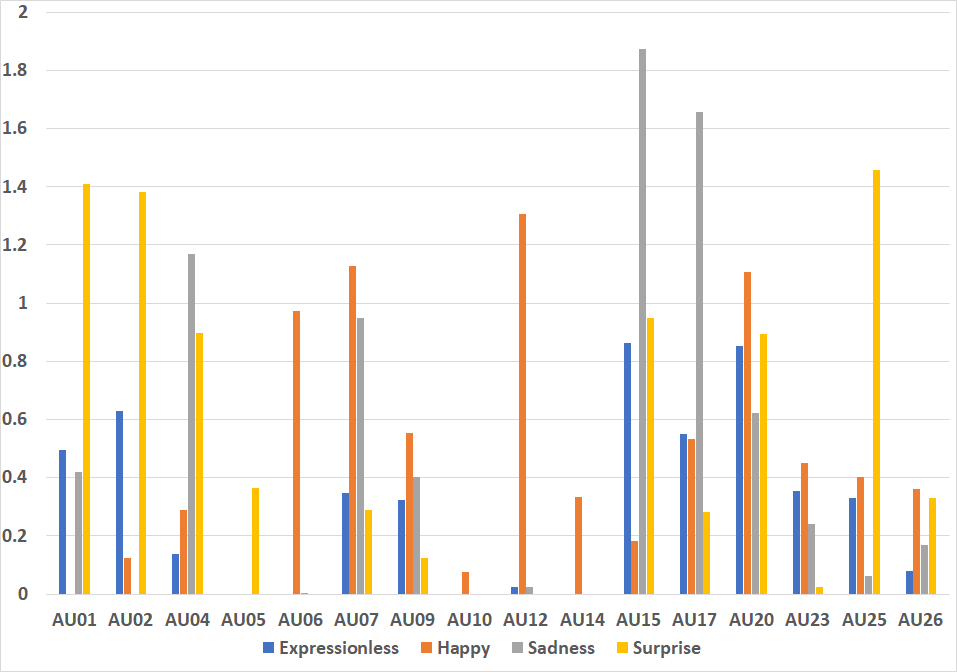}}
		\caption{The AUs value of the four facial expressions: expressionless, happy, sad, surprise.}
	\label{fig:5}
\end{figure}

To get the most representative AUs, we collect a groups of facial images with different distances and angles from the camera for each facial expression, then remove the maximum value and minimum value for each AUs, and average the remaining data, which is shown in Fig. \ref{fig:5}. Furthermore, we select three most responsive AU values for each facial expression as the characteristic vector, e.g., AU06, AU07 and AU12 for happy as shown in Table \ref{tab:my_label}.

\subsection{Human Motion Prediction}

\begin{figure}[!t]
	\centering
	\subfigure[]{
	\includegraphics[width=3.8cm]{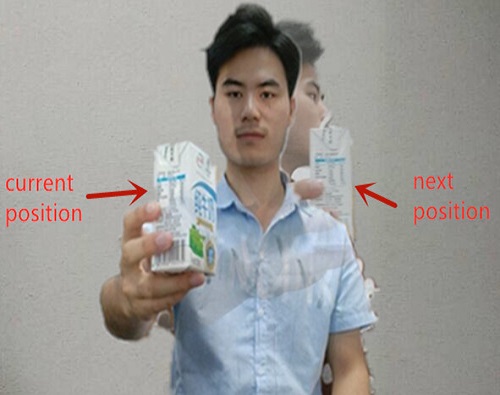}
	}
	\quad
	\subfigure[]{
	\includegraphics[width=3.8cm]{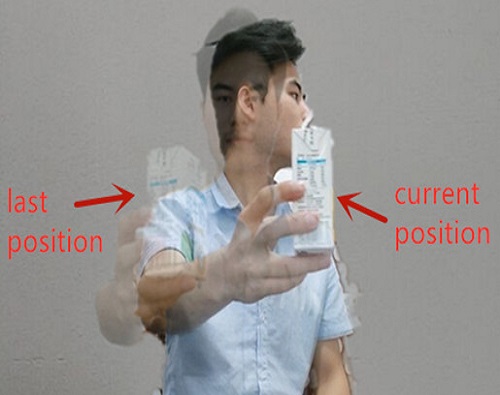}
	}
	\quad
	\subfigure[]{
	\includegraphics[width=3.8cm]{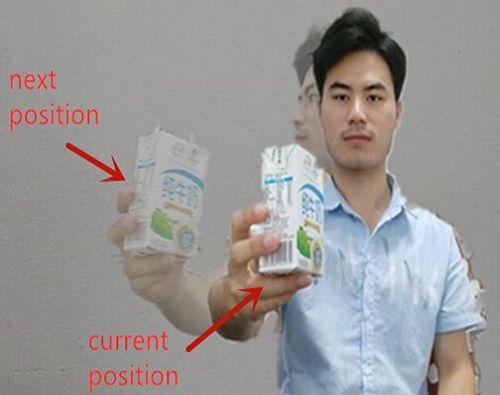}
	}
	\quad
	\subfigure[]{
	\includegraphics[width=3.8cm]{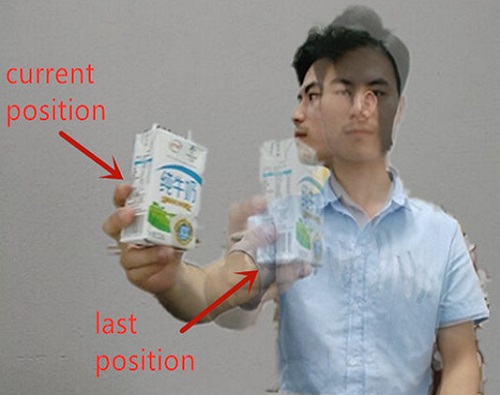}
	}
	\caption{ Head turning can lead to a specific hand motion which can be used for motion prediction. (a) and (b) show the person turns to the left whereas (c) and (d) show the person turns to right. The clear portrait in the image is the current position, whereas the blurred one is the last position or next position. The person is holding a milk box. }
	\label{fig:6}
\end{figure}

Human motion prediction is important for HRI, which can be based on human skeleton detection \cite{butepage2017deep} \cite{gui2018teaching}, human geometric features, deep recurrent neural networks (RNN) to model human motion using probabilistic model \cite{martinez2017human}. However, the detection method based on human skeleton can be unstable as the human body can be occluded or changing rapidly, and neural network depends on the training dataset. The larger the dataset, the better the prediction effect. In contrast, this paper proposes to detect head pose and gaze angle to predict human motion, since our eyes or head will turn to the direction in advance that we plan to move, as shown in Fig. \ref{fig:6}. Therefore, we can detect head pose and gaze angle to predict human movement, this prediction method is simple and effective.

OpenFace 2.0 can not directly provide the head pose and gaze angle, so it is necessary to convert the detection result of OpenFace 2.0 to the real angle of the head and eyes. The transformation from the data of the two pose vectors to real world data (angle of head and eyes turning) can be achieved by:
\begin{align}
    A_t=\frac {(P_t[0]-P_c)}{P_{max}} \frac{\pi}{2}
\end{align}

\begin{itemize}
\item $P_t[0]$ refers to gaze-angle[0] (left-right horizon direction), and $P_c$ means the middle value of gaze-angle[0], $P_{max}$ is the maximum of gaze-angle[0], then we can get true angle of eyes turning $A_t$. 
\item The transformations of head pose and gaze-angle[1] (up-down vertical direction) are similarity to gaze-angle[0], so we just list the transformation of gaze-angle[0].
\end{itemize}

After getting the data of head pose and gaze angle, we need to map the data of head pose and gaze angle to the distance that we plan to move. Taking the scene of people turning to left and right as an example, there are ten volunteers in our experiment. Each volunteer turns randomly to left or right 20 times with different angle, and then we measure the distance of volunteer's arm movement. As for the 20 groups of data of each volunteer, we take the 5 degree as starting point, every 10 degrees as a scale. Afterwards, we can obtain nine scales, which are [5, 15, 25, 35, 45, 55, 65, 75, 85], and then 20 groups of data are divided into different scales according to the angle value. If a certain scale has multiple values, the median value is taken as the value of the scale. The corresponding scale values of ten volunteers eventually are averaged to get the ultimate nine scale values.

\begin{figure}[!t]
	\centerline{
	\includegraphics[width=0.8\linewidth]{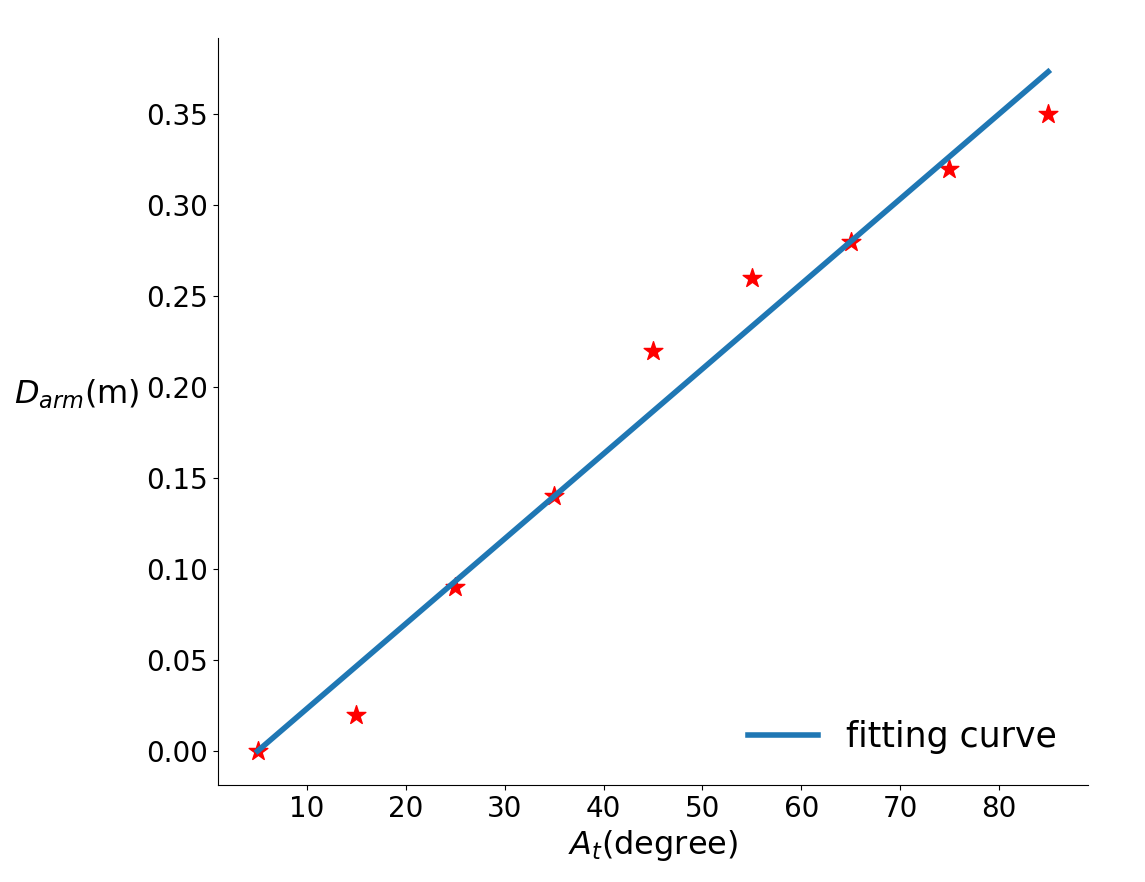}}
		\caption{A linear least square estimation of statistical model between hand movement distance $D_{arm}$ and head turning angle $A_r$. }
	\label{fig:7}
\end{figure}

We then plot  human arm movement distance $D_{arm}$ corresponding to the nine scales of the turning angle $A_t$ on the Fig. \ref{fig:7}, and use a linear ordinary least square estimation to fit the curve, which is 
\begin{align}
    y=\alpha+\beta x
\end{align}
where $\alpha$ and $\beta$ are fitting coefficients to be solved. The fitting curve can be seen in Fig. \ref{fig:7}, such that we can estimate the move distance according to different head pose and gaze angle.

\section{Psychological safety and motion prediction for motion planning}

In this paper, safeguarding human psychological safety is achieved by the control of robot's velocity and trajectory. When robot's velocity and trajectory change greatly, it can cause discomfort or stress to the human. Therefore, it is necessary to constrain robot's velocity and trajectory to ensure human psychological safety. Based on our previous SDAPF algorithm for motion planning, we here introduce an improved version P-SDAPF to consider psychological safety according to the human facial expression and attention changes. Therefore, here we first introduce the SDAPF briefly, and then introduce the P-SDAPF to combine the psychological safety.

\subsection{Sampling and Danger-Index Based Artificial Potential Field (SDAPF)}

Our previous work introduces a sampling and danger-index based artificial potential field (SDAPF) \cite{liu2020online} algorithm to optimize motion planning. Compared to traditional APF, SDAPF introduces the speed repulsive force function $F_{rev}$ with improved impact factors and adaptive adjustment of step-size $S$ to overcome the shortcomings of APF. The impact factors include that the distance impact and the speed impact factor.

The distance impact factor is affected by the distance between the robot and dynamic obstacle. The closer the distance, the greater the distance factor. The distance influence factor function is:
\begin{equation}
	f_d=\left\{
	\begin{aligned}
	\eta (\frac{1}{p(X)}-\frac{1}{p_{max}})  & , \ p(X) \leq p_{max} \\
	0 \qquad  & , \ otherwise
	\end{aligned}
	\right.
	\label{eq:f_d}
\end{equation}
The $\eta$ is the distance scale factor, which can be expressed as:
\begin{align}
    \eta=\frac{p_{max} p_{min}}{p_{max} - p_{min}}
\end{align}
where $p_{max}$ , $p_{min}$ are the max and min influence radius of the moving obstacle, and $p(X)$ represents the distance between the robot and the obstacle at position X.

In the same way, the speed influence factor depends on the relative speed between the robot and dynamic obstacle, which can be expressed as:
\begin{align}
    k_v=sgn(\gamma |v_o|-|v_r|)
    \label{eq:k_v}
\end{align}
where $|v_o|$ is the magnitude of velocity of the dynamic obstacle, and $|v_r|$ means the magnitude of the robot's velocity. The $\gamma$ is just a speed scale factor. The $sgn()$ is the sign function, which makes $k_v$ being a positive integer when $\gamma$ times $|v_o|$ is greater than $|v_r|$ and vice versa. The $k_v > 0$ means dynamic obstacle moving fast, which influences the velocity repulsive force. Then SDAPF can adaptively adopt dynamic obstacle avoidance strategy for the situation.

The expression of the velocity repulsive force is:

\begin{equation}
F_{rev}(v)=\left\{
\begin{aligned}
k_{ro}f_d(v_r-v_o) & ,\ k_v \leq 0 \cap f_d > 0 \cap \alpha \in (-\frac{\pi}{2},\frac{\pi}{2}) \\
k_{ro}f_d(v_r+v_o) & ,\ k_v > 0 \cap f_d > 0 \cap \alpha \in (-\frac{\pi}{2},\frac{\pi}{2}) \\
0 \qquad  & , \qquad otherwise
\end{aligned}
\right.
\label{equ:f_rev}
\end{equation}
where $k_{ro}$ is the scale factor, $k_v$ is the speed influence factor, $f_d$ is distance influence factor, $v_r$ is the velocity of end-effector, $v_o$ is the velocity of obstacle, $\alpha$ is the angel between the relative velocity vector $v_{or} = v_r - v_o$ and the displacement vector from the end-effector to the obstacle. Whether $k_v$ is greater than 0 only affects the addition and subtraction relationship between $v_r$ and $v_o$. When $k_v$ is less than or equal 0, $v_r$ subtracts $v_o$. In contrast, $v_r$ plus $v_o$, which can be seen as the relative velocity between the robot and dynamic obstacle moving in the opposite direction at -$v_o$.

\begin{equation}
S=\left\{
\begin{aligned}
& v_{r} \bigtriangleup t_m,\ P(X_{t},X_{ot}) > p_d \cap  \bigtriangledown P(X_{t},X_{ot}) \leq 0 \\
& d_{0},\quad \quad \ \ P(X_t,X_{ot}) \leq p_d \\
& toGoal, \ \ P(X_t,X_{ot}) > p_d \cap \bigtriangledown P(X_t,X_{ot}) > 0
\end{aligned}
\right.
\end{equation}
where $d_{0}$ represents the initial step size, $v_{r}$ is the velocity of end-effector, $t_{m}$ is the maximum time that the manipulator can move when the distance between the end-effector and the obstacle is greater than $p_{d}$, which is a parameter that should be guaranteed to be greater than or equal to the max influence radius of the moving obstacle. $P(X_{t},X_{ot})$ is the distance between robot's end-effector and obstacle. $\bigtriangledown$ $P(X_{t},X_{ot})$ is the derivative of $P(X_{t},X_{ot})$. $toGoal$ represents that we can directly select the goal point as the next position when the end-effector is outside the influence range of the obstacle and the relative distance is increasing.

\subsection{Adaptive Adjustment of Velocity and Step-size Based on Facial Expression Recognition}

An adaptively adjustment method of velocity and step-size according to facial expressions, head pose and gaze angle is required to ensure psychological safety in the dynamic environment. When the expression detected is unusual or head and eyes turn to other direction, robot's velocity and step-size need to be reduced, which can be defined as:
 
\begin{equation}
    \begin{aligned}
        V_p=V_t+(P_t-P_{t-1})V_{max}A_h/ \frac{\pi}{2}\\
    \end{aligned}
\end{equation}
\begin{equation}
    \begin{aligned}
        V_{t+1}=(1-H) (V_t-E_x V_b)+ H V_p \\
    \end{aligned}
\end{equation}

\begin{itemize}

\item The setting function of step-size is similarity to the function of velocity, so we only list and explain the function of velocity.

\item $V_t$ refers to the velocity of robot at time $t$, which is a 3D vector including X, Y, Z axis. $V_{max}$ is the maximum of velocity that robot can increase or decrease at the next moment. $A_h$ is angle of head and eyes turning. $P_t$ is human head and gaze pose at time $t$ in HRI, so $P_t - P_{t-1}$ represents the orientation of head and gaze will turn. It works only when the distance between human and robot is small. To sum these two parts, we can get robot's velocity $V_p$ affected by head and gaze pose. $V_b$ is the change range of the foundation, and $E_x$ is a flag of whether the expressions is normal or not, i.e., the value is $0$ when facial expressions are normal and vice versa. $H$ is a flag of whether head and eyes turn or not, which is $1$ when head and eyes turn more than a certain angle, otherwise it is $0$.
\end{itemize}

To control the step size of end-effector according to the state of the human, we set the initial step size as the maximum value, and decrease the step size each period if abnormal facial expression is detected, or the head pose and gaze angle exceed a threshold, until it reaches the minimum value. When the facial expression, head pose and gaze angle is normal, we increase the step size by each period until it reaches the maximum value.  
 
\subsection{Motion Prediction Based on Head Pose and Gaze Angle Estimation for Trajectory Optimization}

Human takes action when head and gaze change, such that robot's original trajectory needs to be optimized to ensure human safety. The net force of SDAPF is given by 
\begin{align}
    F=F_{att}+F_{rep}+F_{rev}
\end{align}
\begin{equation}
\left\{
\begin{aligned}
& F_{rep1}=k_r(\frac{1}{p(X)}-\frac{1}{p_0})\frac{1}{p^2(X)}(X-X_g)^n \\
& F_{rep2}=-n \frac{k_r}{2}(\frac{1}{p(X)}-\frac{1}{p_0})^2(X-X_g)^{n-1} \\
& F_{rep}=F_{rep1}+F_{rep2}
\end{aligned}
\right.
\end{equation}
where $n$ is a constant that is greater than zero, $k_r$ is the scale factor, $(X-X_g)$ represents the distance between the robot and target, $p(X)$ represents the distance between the end-effector and the obstacle at position $X$, and $p_0$ is the influence radius of each obstacle.

We optimize $F_{rep}$ and $F_{rev}$ by using head pose and gaze angle to predict human motion. For $F_{rep}$ and $F_{rev}$,  we use predicted human pose $P_{pre}$ of equation \ref{eq:p_pre} and predicted velocity $V_{pre}$ of equation \ref{eq:v_pre} to replace $p(X)$ of equation \ref{eq:f_d} and $v_o$ of equation \ref{eq:k_v} respectively, which are defined as:
\begin{align}
    P_{pre}=P_{ot}+(P_{t}-P_{t-1})P_{max}A_h/ \frac{\pi}{2}
    \label{eq:p_pre}
\end{align}
\begin{align}
    V_{pre}=\frac{(P_{pre}-P_{ot})}{\bigtriangleup t}
    \label{eq:v_pre}
\end{align}
where $P_{ot}$ is the human arm pose at present moment, and ($P_{t}-P_{t-1}$) represents the orientation of head and gaze will turn as equation (4). $P_{max}$ is the maximum amplitude of human arm move when head and gaze turn $\frac{\pi}{2}$, and $A_h$ is the angle of head and gaze turning. $\bigtriangleup t$ is the cycle of program execution.

\section{Experiment and analysis}

This paper proposes a motion planning algorithm that combines psychological safety and motion prediction. Here psychological safety is mainly affected by the robot's velocity and trajectory, so our experiments analyse velocity and trajectory of robot while facial behavior is changing. To demonstrate the proposed idea, we first use the Gazebo to build a simulated HRI environment to show how the robot can capture human emotion changes and predict the motion of human hand, and then a real TIAGo robot is used to show the practical performance of the proposed P-SDAPF method. 

\subsection{Dynamic Interaction Experiment in Gazebo Environment}

In order to facilitate the experiment, we choose Gazebo to simulate dynamic interaction environment. In Gazebo, the TIAGo robot need to avoid the moving hand of another robot (simulated human hand) to pick a box from table. In order to track dynamic motion of the simulated human hand precisely, we use an ArUco Marker detection method. The ArUco markers are attached to the simulated human hand and box, such that the robot can estimate the relative pose using these visual markers. The human facial expression, head pose and gaze angle are estimated by OpenFace 2.0 using a real camera. 

\begin{figure}[!t]
\centering
\subfigure[3D trajectories by SDAPF]{
\includegraphics[width=3.8cm]{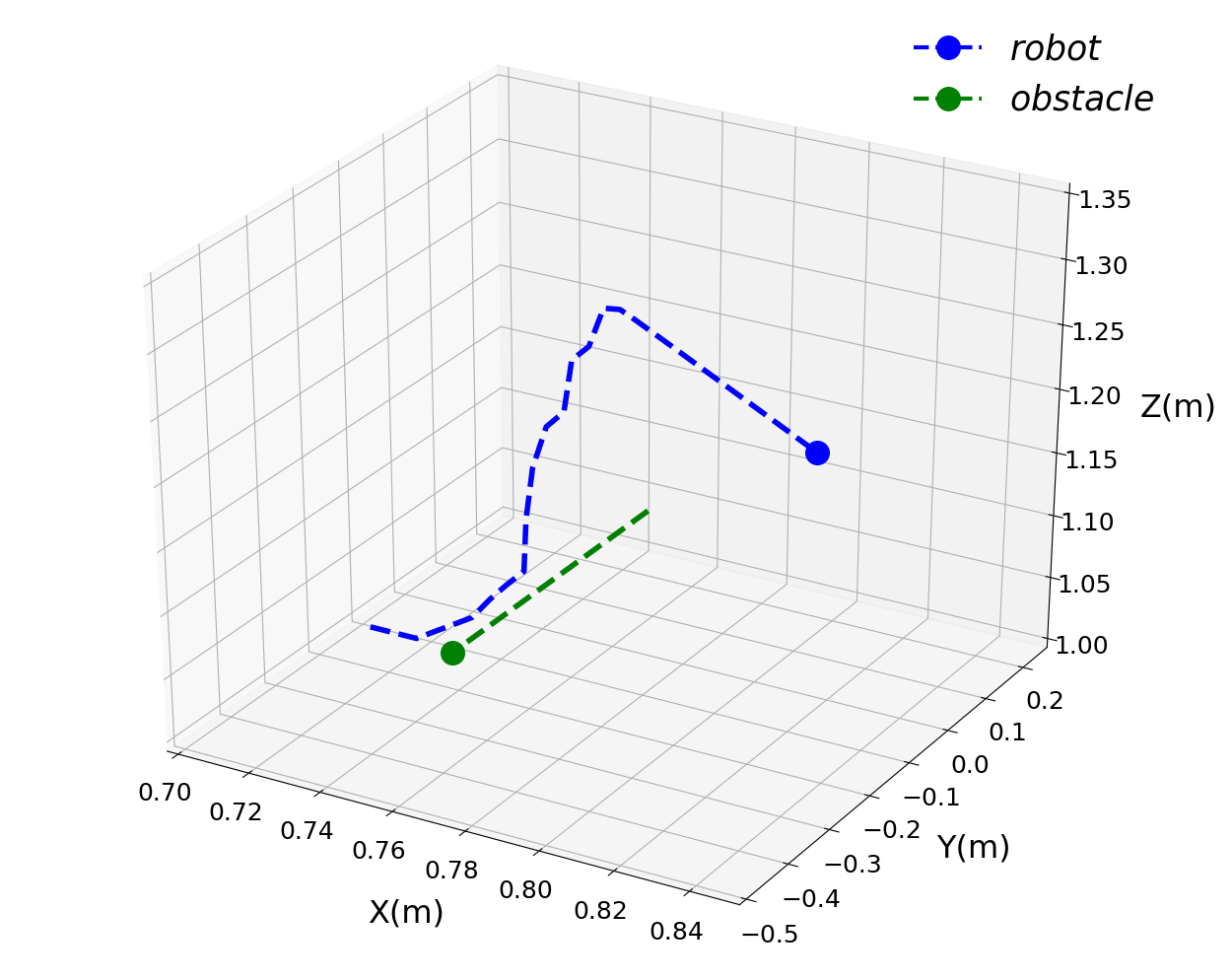}
}
\quad
\subfigure[3D trajectories by ours]{
\includegraphics[width=3.8cm]{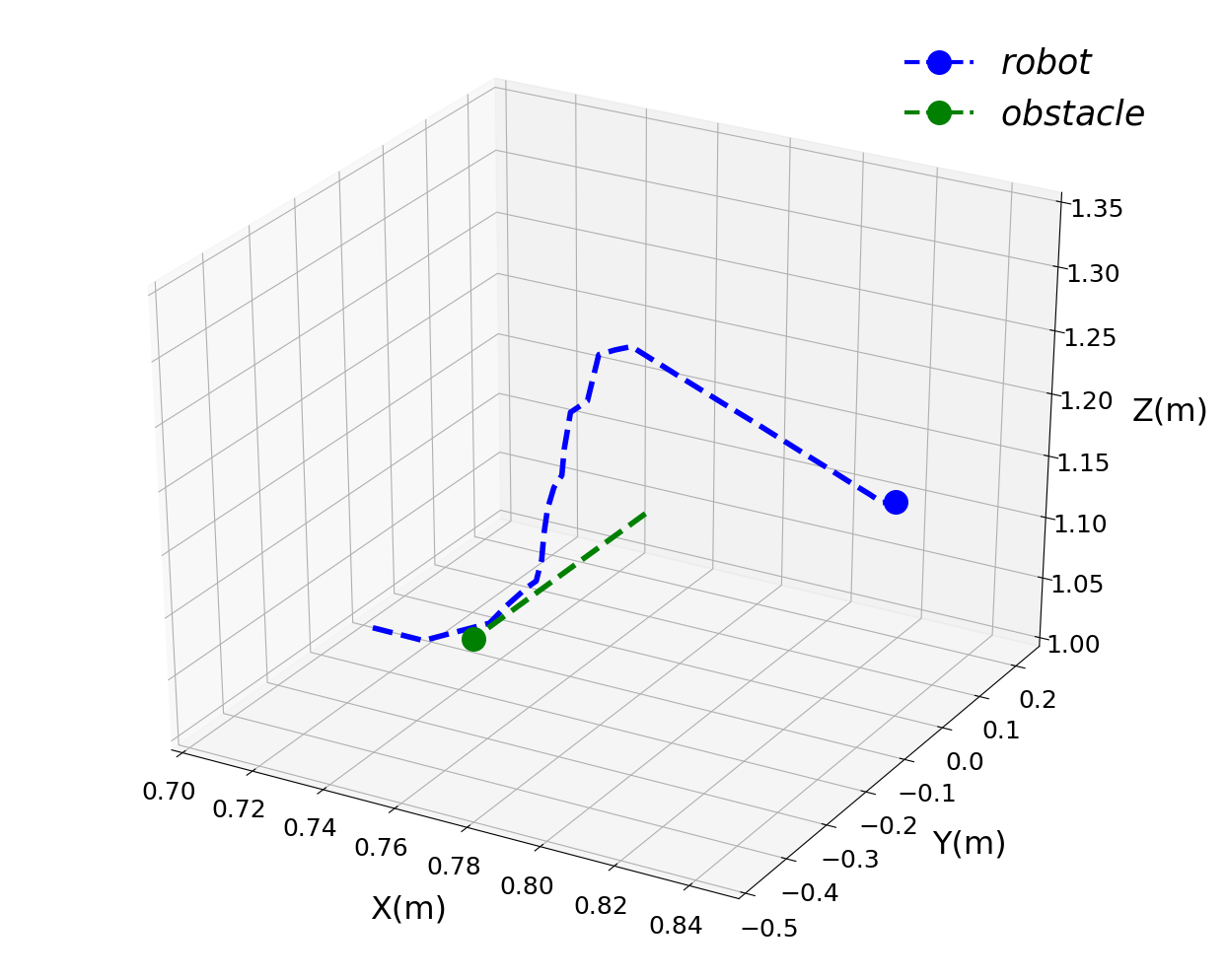}
}
\quad
\subfigure[2D trajectories, step size, velocity and facial expression by SDAPF]{
\includegraphics[width=3.8cm]{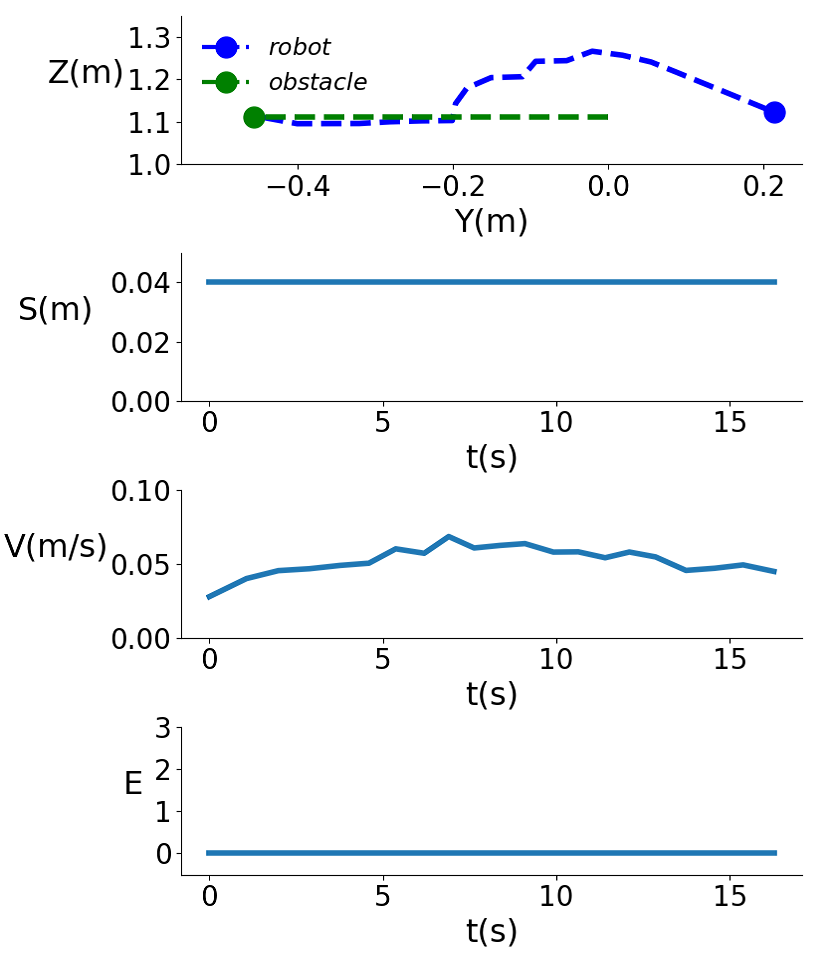}
}
\quad
\subfigure[2D trajectories, step size, velocity and facial expression by P-SDAPF]{
\includegraphics[width=3.8cm]{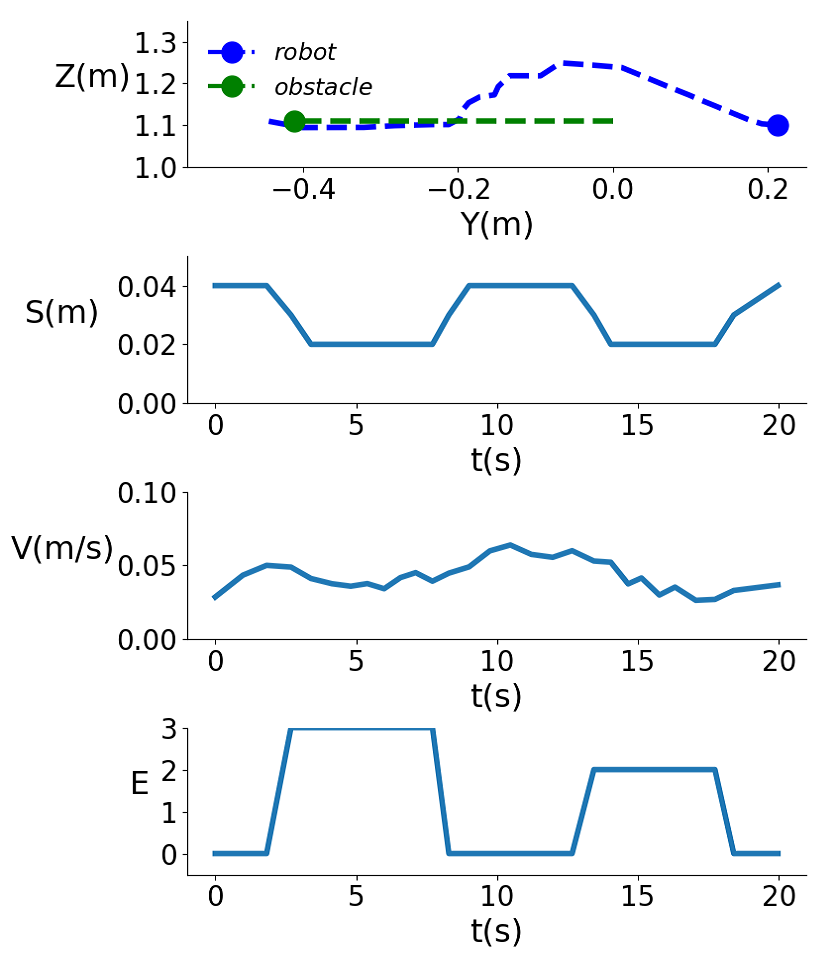}
}
\quad
\subfigure[Happy]{
\includegraphics[width=3.8cm]{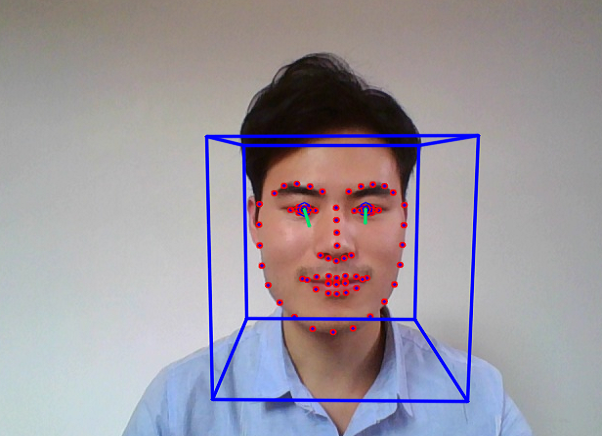}
}
\quad
\subfigure[Sad]{
\includegraphics[width=3.8cm]{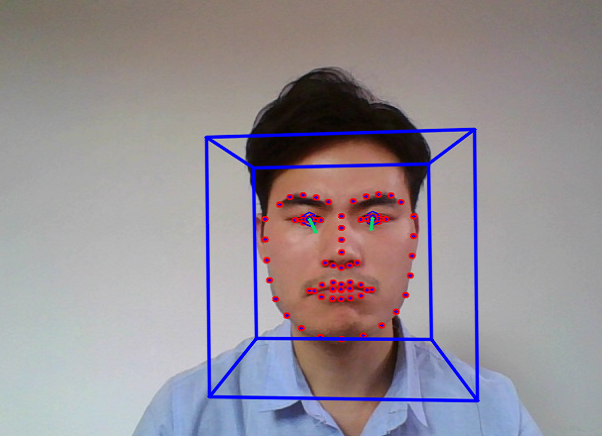}
}
\caption{Motion planning considering facial expression of human: (a) and (c) are control results of the original SDAPF, whereas (b) and (d) are results of the proposed P-SDAPF method by considering facial expressions, (e) and (f) are sample images of different facial expressions which correspond to happy and sad respectively. (a) and (b) are 3D trajectories of robot and moving obstacle (human hand). (c) and (d) show the projected 2D trajectories on YZ plane, step size $S$, velocity $V$, facial expressions $E$.}
\label{fig:exp-facial}
\end{figure}

Firstly, we explore the influence of facial expressions on TIAGo's trajectory, velocity and step size. As shown in Fig.\ref{fig:exp-facial}, when the change of facial expression is detected (t=2s, 13s), TIAGo's velocity becomes slower and step size becomes shorter (the lower limit is 0.02) as shown in Fig.\ref{fig:exp-facial} (b) and (d) compared to the original APF algorithm without psychology consideration as shown in Fig.\ref{fig:exp-facial} (a) and (c), which helps to ensure human psychological safety. In Fig.\ref{fig:exp-facial}, the figures (a) and (c) are 3D trajectories of gripper (blue) and simulated human hand (green), the figures (b) and (d) show projected trajectories on YZ plane, step size, velocities and detected facial expressions (i.e., 0 means expressionless, 1 means surprise, 2 means sadness and 3 means happiness). The subfigures (e) and (f) are sampled facial images that shows the happiness and sadness. The TIAGo's velocity and step size return to normal values when detected facial expressions is normal. From the experimental results, the TIAGo's trajectory and velocity that combines facial expression recognition are more smooth and comfortable than the one without facial expression recognition which can cause stress and danger for human.

\begin{figure}[!t]
\centering
\subfigure[3D trajectories by SDAPF]{
\includegraphics[width=3.8cm]{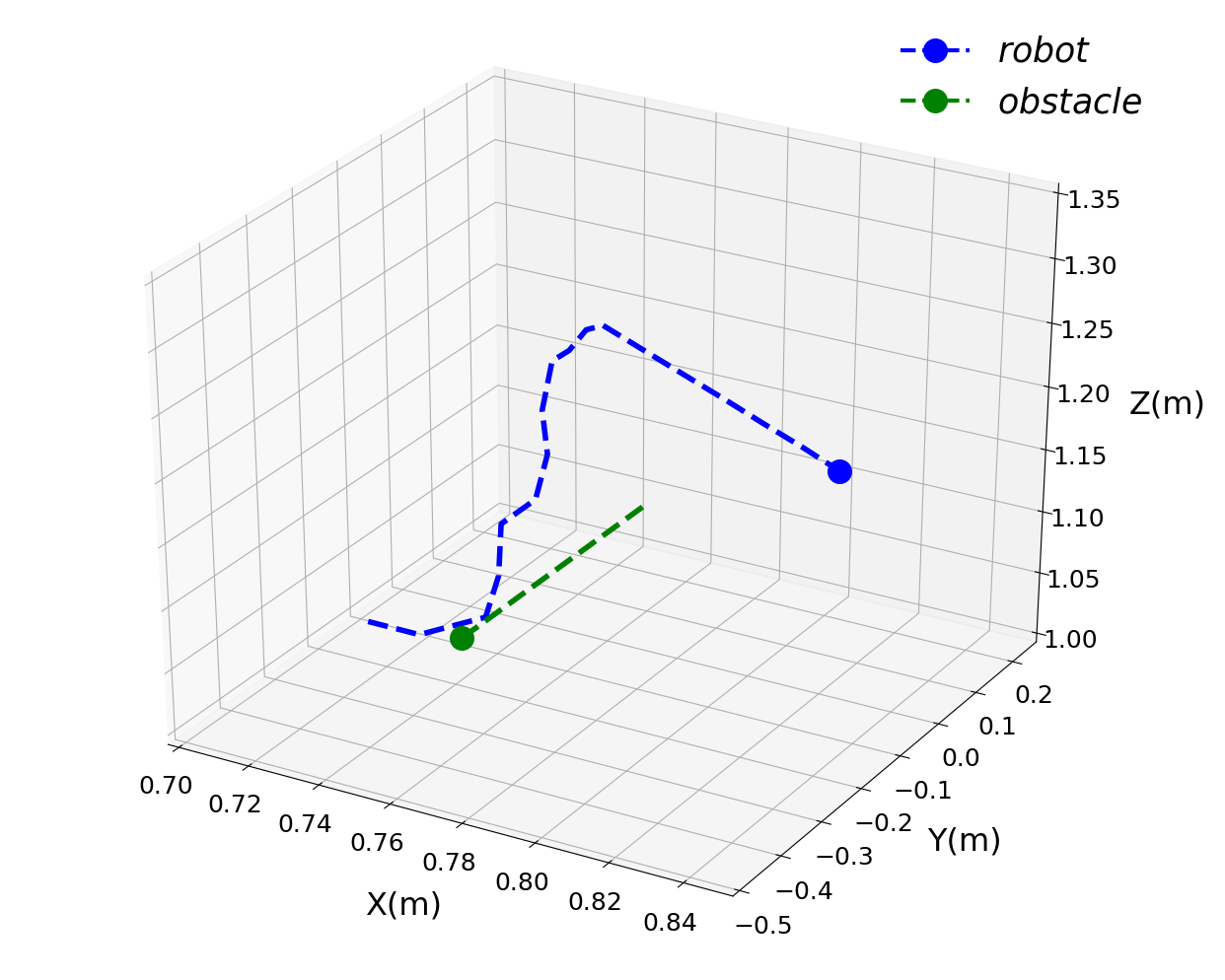}
}
\quad
\subfigure[3D trajectories by ours]{
\includegraphics[width=3.8cm]{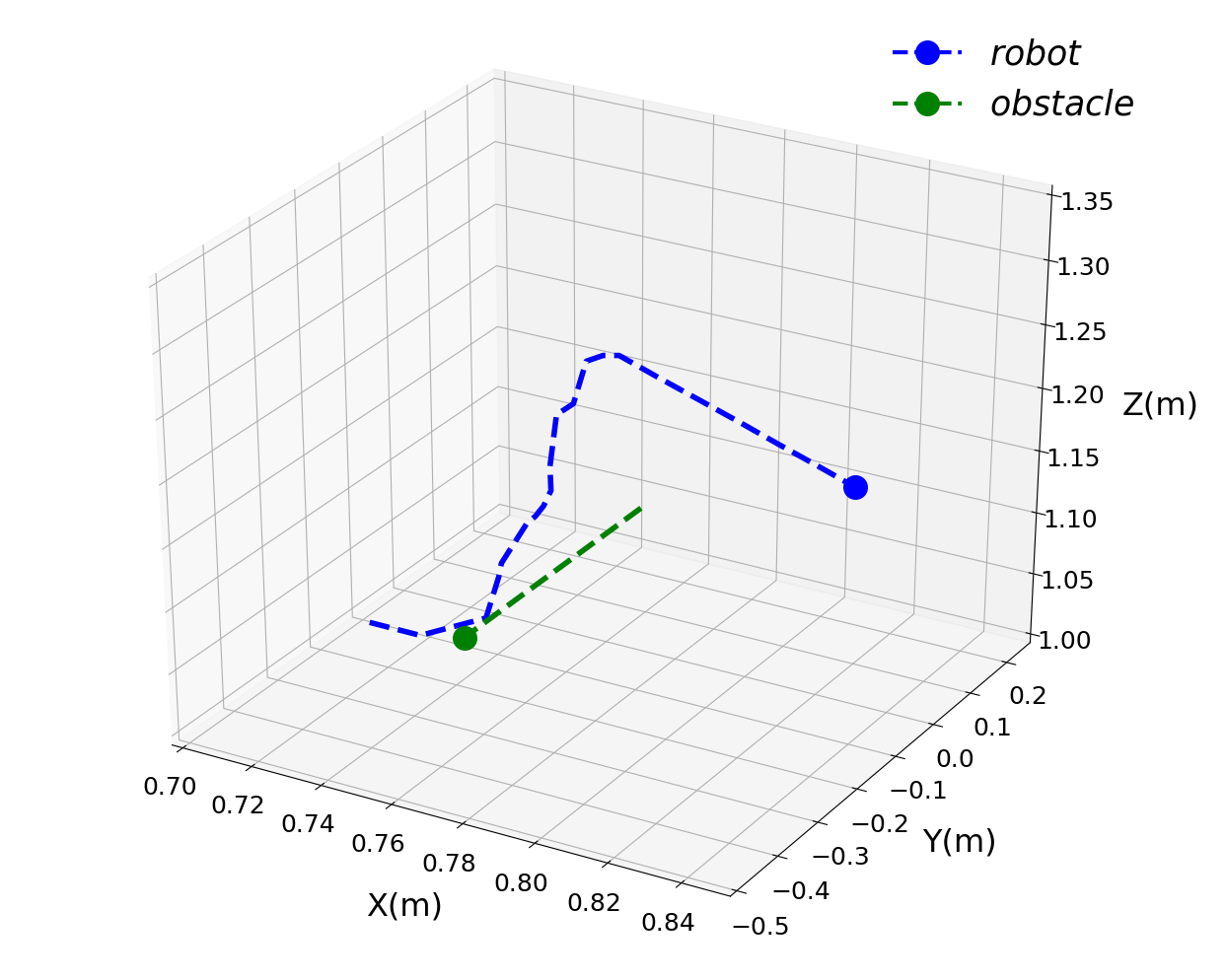}
}
\quad
\subfigure[2D trajectories, step size, velocity and  turning angle of head and gaze by SDAPF]{
\includegraphics[width=3.8cm]{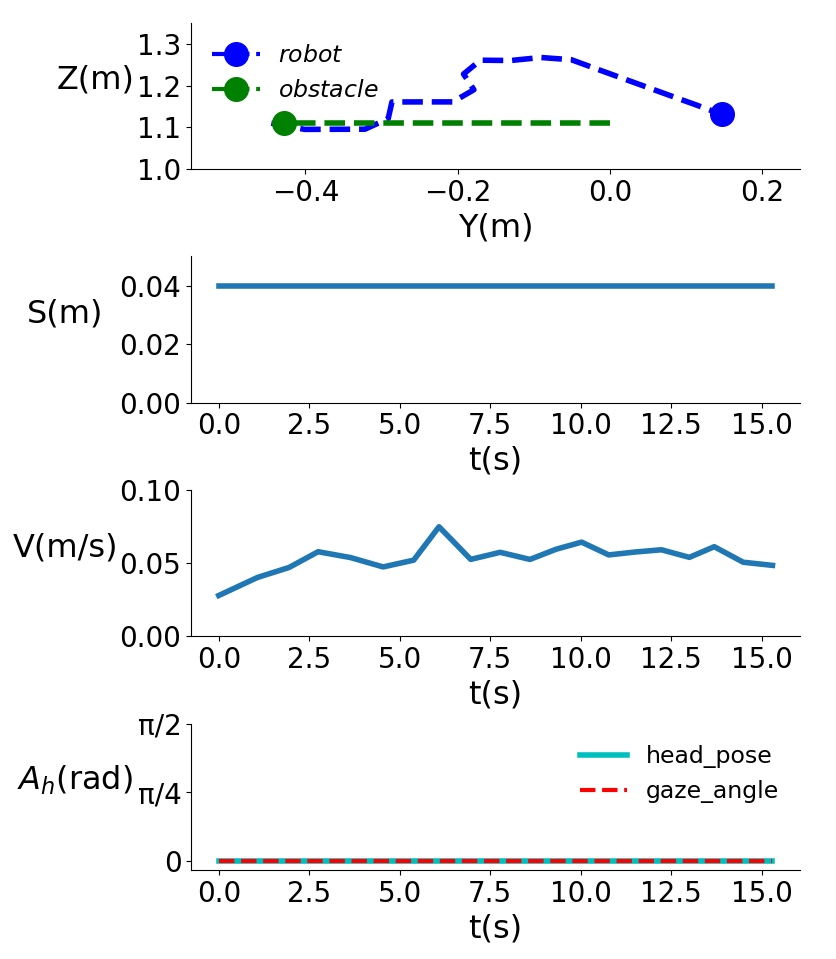}
}
\quad
\subfigure[2D trajectories, step size, velocity, turning angle of head and gaze by P-SDAPF]{
\includegraphics[width=3.8cm]{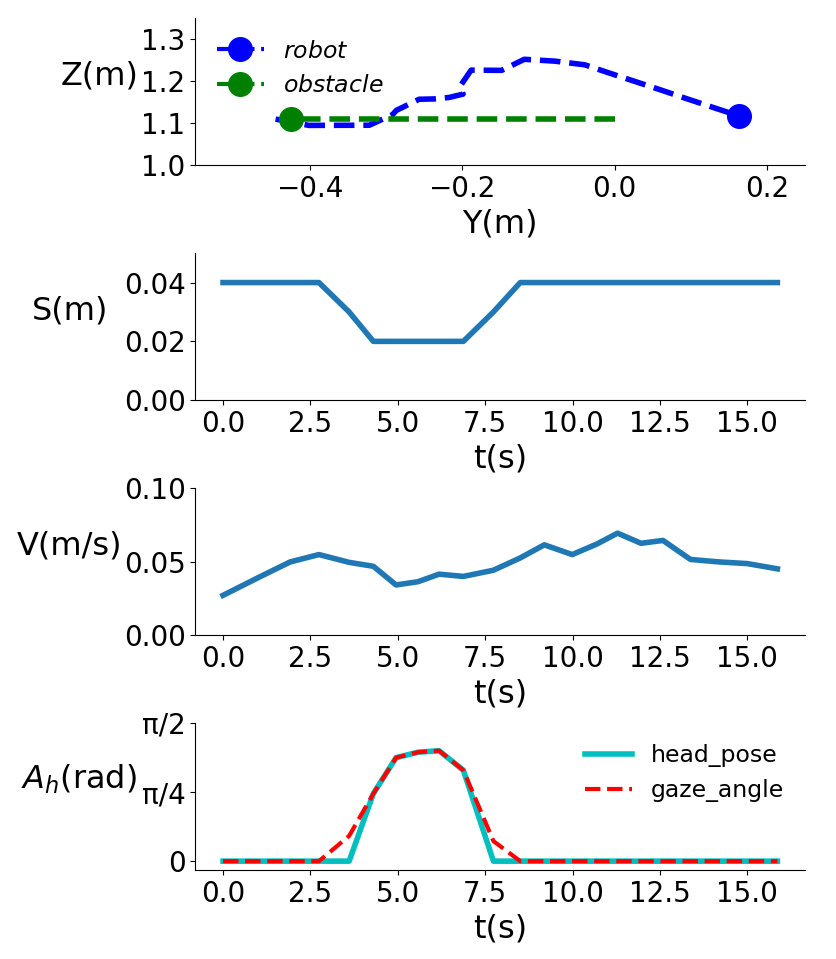}
}
\quad
\subfigure[Turning direction]{
\includegraphics[width=3.8cm]{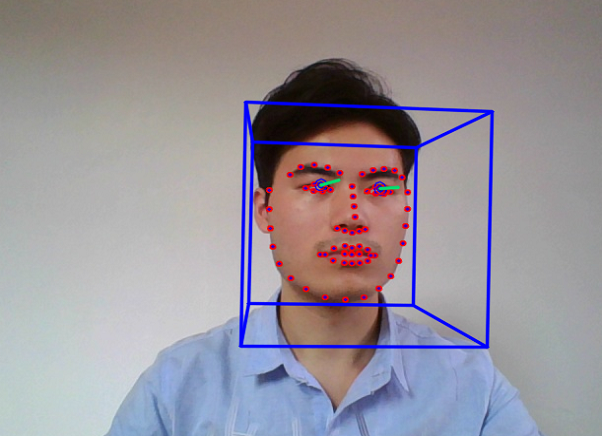}
}
\quad
\subfigure[Turning direction]{
\includegraphics[width=3.8cm]{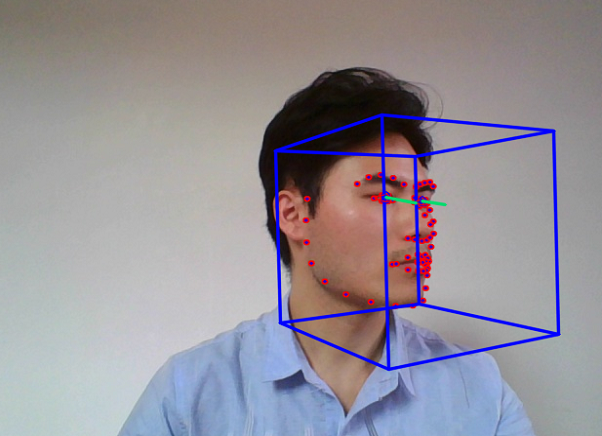}
}
\caption{Motion planning considering head pose and gaze angle of human: (a) and (c) are control results of the original SDAPF, whereas (b) and (d) are results of the proposed P-SDAPF method by considering head pose and gaze angle, (e) and (f) are sample images of different head pose and gaze angle. }
\label{fig:exp-head}
\end{figure}

The ability to predict the moving direction of human can improve the performance of path planner of the TIAGo robot. As shown in Fig.\ref{fig:exp-head}, without the human motion prediction, the projected trajectory on YZ plane in Fig.\ref{fig:exp-head} (c) has several sharp turning points due to the dynamic obstacle (human hand), whereas our method can improve such situation by predicting the human motion using the human head pose and gaze direction information as shown in Fig.\ref{fig:exp-head} (d). At t=2.5s, TIAGo's velocity slows down and the step size is decreased when the head and gaze are turning. The predicted trajectory of dynamic human hand helps TIAGo to optimize trajectory to avoid dynamic obstacle effectively. In our experiment, when the turning angle of human head is less than $22$ degrees, we only consider the gaze angle and set head angle to $0$. Otherwise we only consider the head angle and set the gaze angle to the head angle as shown in the last curve of Fig.\ref{fig:exp-head} (d). Fig.\ref{fig:exp-head} (e) and (f) are the sample facial images of turning around.  

\subsection{Dynamic Interaction Experiment in Real Environment}

\begin{figure}[!t]
	\centering
	\subfigure[3D trajectories by SDAPF]{
	\includegraphics[width=3.8cm]{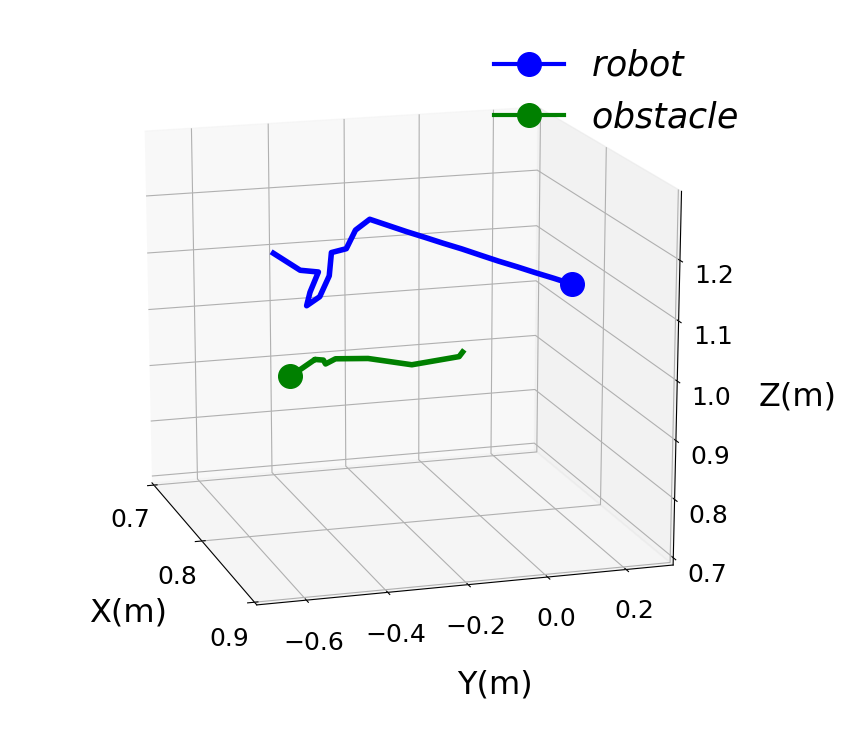}
	}
	\quad
	\subfigure[3D trajectories by ours]{
	\includegraphics[width=3.8cm]{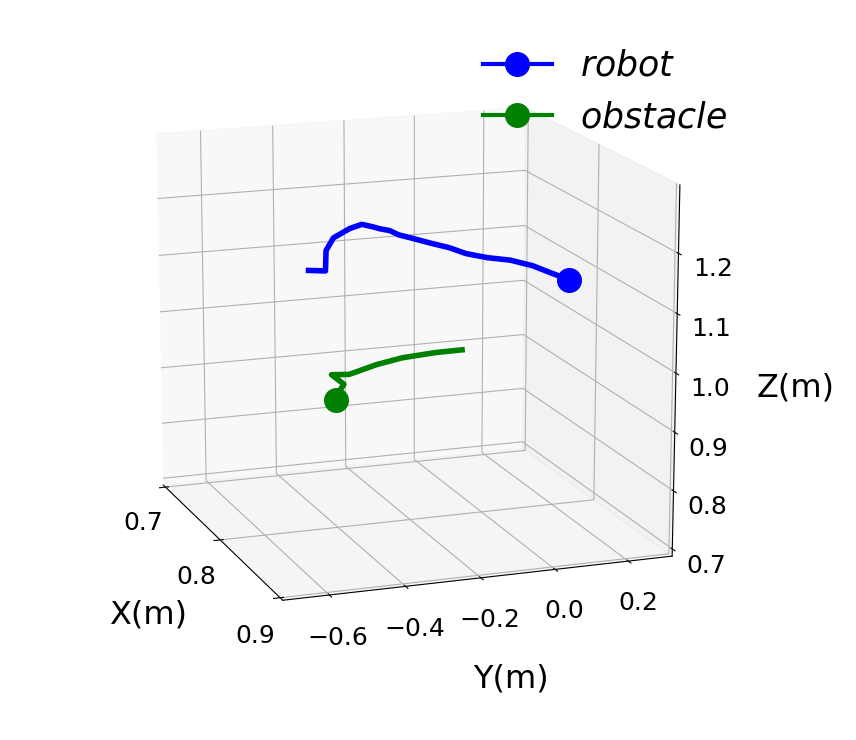}
	}
	\quad
	\subfigure[2D trajectories, step size, velocity, turning angle of head and gaze, and facial expression by SDAPF]{
	\includegraphics[width=3.8cm]{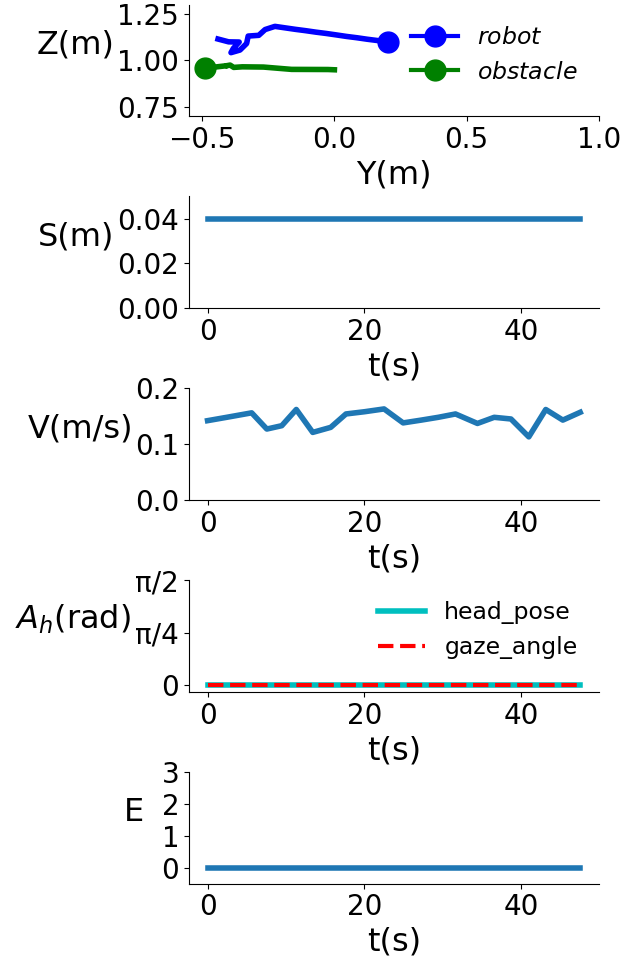}
	}
	\quad
	\subfigure[2D trajectories, step size, velocity, turning angle of head and gaze, and facial expression by P-SDAPF]{
	\includegraphics[width=3.8cm]{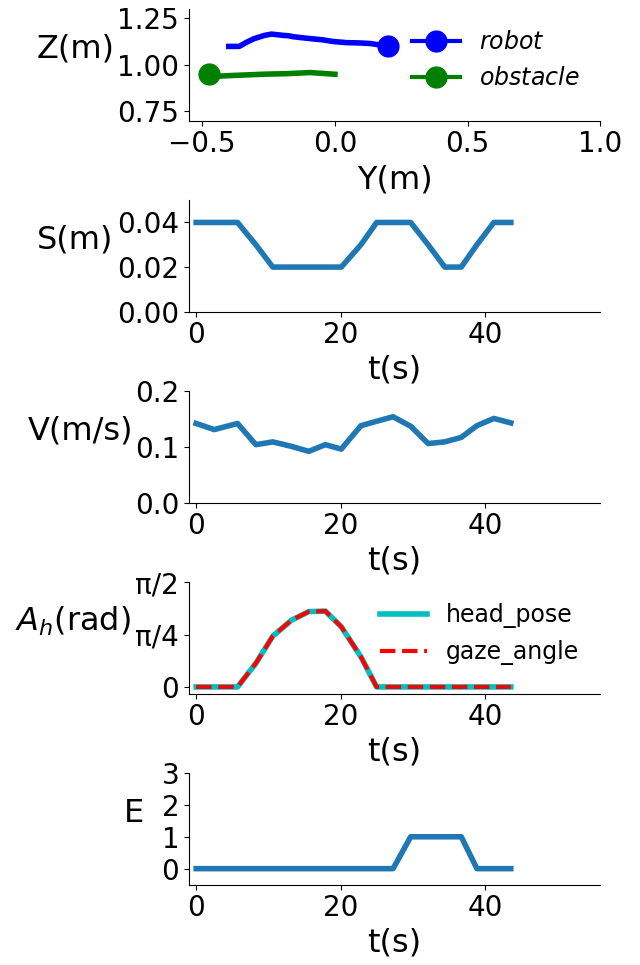}
	}
	\caption{Experimental results of dynamic obstacle avoidance using SDAPF and our P-SDAPF respectively. The trajectory using SDAPF has sharp turning points and fall back, while our P-SDAPF has a smoother trajectory due to its motion prediction ability.}
	\label{fig:real-curve}
\end{figure}

\begin{figure*}[!t]
	\centering
	\subfigure{
	\includegraphics[width=4.0cm,height=3cm]{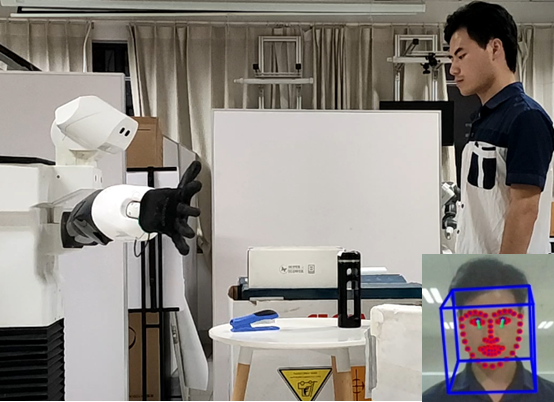}
	}
	\
	\subfigure{
	\includegraphics[width=4.0cm,height=3cm]{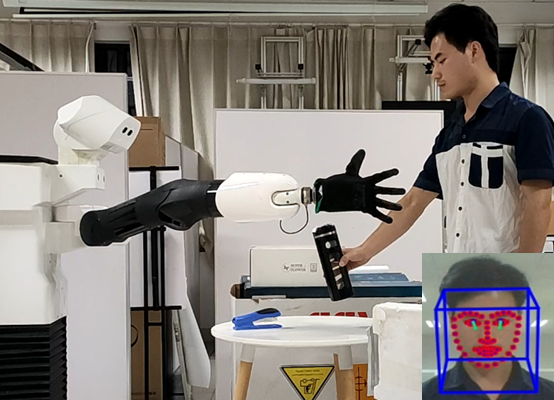}
	}
	\
	\subfigure{
	\includegraphics[width=4.0cm,height=3cm]{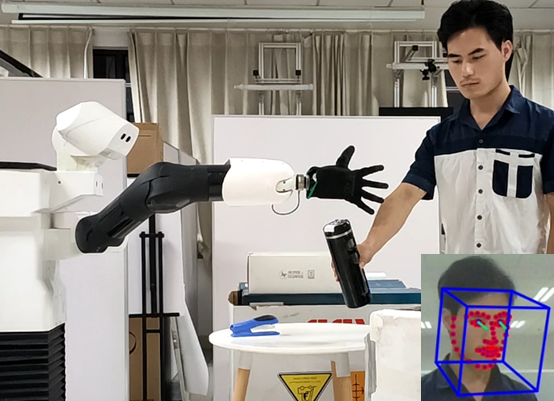}
	}
	\
	\subfigure{
	\includegraphics[width=4.0cm,height=3cm]{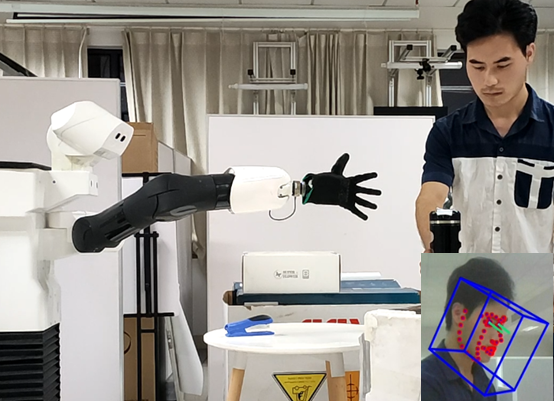}
	}
	\
	\subfigure{
	\includegraphics[width=4.0cm,height=3cm]{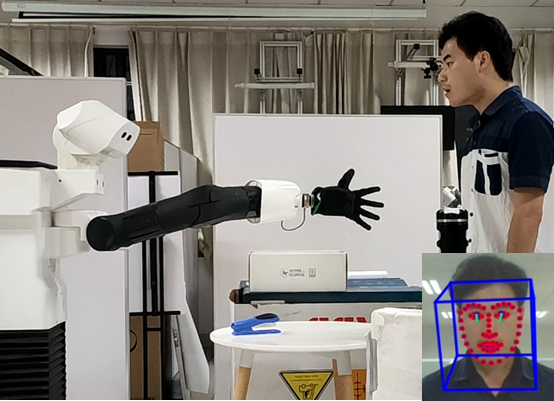}
	}
	\
	\subfigure{
	\includegraphics[width=4.0cm,height=3cm]{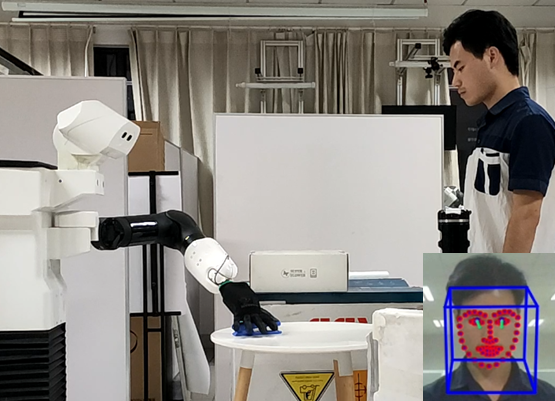}
	}
	\
	\subfigure{
	\includegraphics[width=4.0cm,height=3cm]{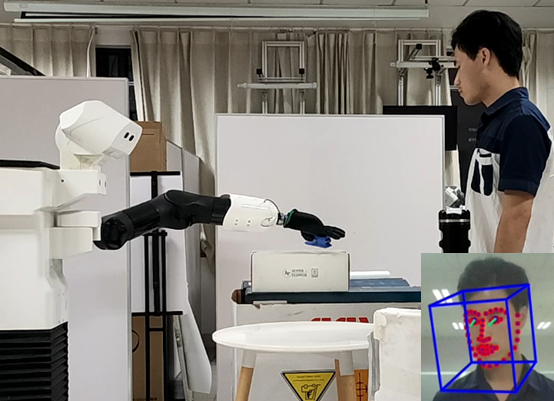}
	}
	\
	\subfigure{
	\includegraphics[width=4.0cm,height=3cm]{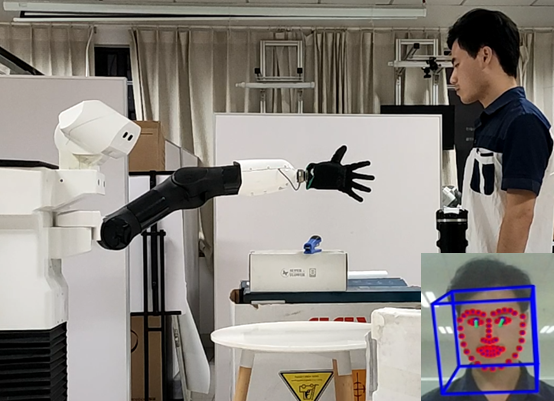}
	}
	\caption{The captured image sequences for dynamic obstacle avoidance using our P-SDAPF algorithm.}
	\label{fig:real-P-SDAPF}
\end{figure*}

To verify the effectiveness and applicability of our approach, we use a real dynamic interaction scene with a TIAGo robot as shown in Fig. \ref{fig:real-P-SDAPF}. TIAGo needs to grab the stapler from the table, and then place the stapler at the right side of the human. At the same time, the human picks up the water cup on the table and place it at the left side of human. In the process of this scenario, there have overlapping operation workspaces between human arm and TIAGo's end-effector. TIAGo needs to recognize the pose of water cup and human arm, and then avoid possible collisions during the interaction. 

The numerical results can be seen in Fig. \ref{fig:real-curve}, which shows the 3D trajectory, 2D trajectory, step size, velocity, turning angle of head or gaze and facial expression using SDAPF and our P-SDAPF methods respectively.  The trajectory of the end-effector using SDAPF has several sharp turning points and fall back due to the sudden movement of human arm. In contrast, our P-SDAPF method can achieve a smoother trajectory since we can predict the human motion by detecting the human head pose and gaze angle. Meanwhile, we also show that the TIAGo's velocity becomes slower and step size becomes shorter when facial expression, head pose and gaze angle change, which help to ensure human physical and psychological safety. Fig. \ref{fig:real-P-SDAPF} show the captured images of demonstration using two different algorithms.

\section{Conclusion}

In this paper, we propose a motion planning algorithm combining psychological safety and motion prediction for a sense motive robot. Safety is a core problem of human robot interaction and collaboration, which includes not only physical safety, but also psychological safety. Our method aims to solve safety problems by optimizing velocity control and step size adjustment method using facial expression information, head pose and gaze angle estimation in a sampling based APF scheme. From the experimental results using a 7-DOF TIAGo robot in the 3D Gazebo environment and real HRI environment, we show that our robot can recognize the human emotion state and predict human motion, such that it can control velocity and step size accordingly. In this way, both the physical safety and psychological safety can be ensured, and the interaction experience can be improved. In future, we would like to introduce more safe factors into our framework to reinforce psychological safety and physical safety of HRIC, e.g., social customs, speech interaction.

\bibliographystyle{ieeetr}
\bibliography{WF}





\end{document}